\documentclass[pdflatex,sn-mathphys-ay]{sn-jnl}

\usepackage{graphicx}%
\usepackage{multirow}%
\usepackage{amsmath,amssymb,amsfonts}%
\usepackage{amsthm}%
\usepackage{mathrsfs}%
\usepackage[title]{appendix}%
\usepackage{xcolor}%
\usepackage{textcomp}%
\usepackage{manyfoot}%
\usepackage{booktabs}%
\usepackage{algorithm}%
\usepackage{algorithmicx}%
\usepackage{algpseudocode}%
\usepackage{listings}%
\usepackage{subcaption} 

\theoremstyle{thmstyleone}%
\theoremstyle{thmstyletwo}%

\theoremstyle{thmstylethree}%

\begin{document}

\title[Article Title]{Coupled Modeling of External Pressure and Capillary Blood Flow: Nonlinear Dynamics of Vascular Elasticity and Collapse Effects}


\author[1]{\fnm{Xinyi} \sur{Cai}}\email{2984266909@qq.com}
\author[2]{\fnm{Hanyue} \sur{Mo}}\email{mohanyue808@gmail.com}
\author[3]{\fnm{Yuyao} \sur{Chang}}\email{3206636993@qq.com}
\author*[2]{\fnm{Kun} \sur{Cheng}}\email{chengkun@ustc.edu}

\affil[1]{\orgdiv{College of Life Sciences}, \orgname{China Jiliang University}, \orgaddress{\street{No. 258 Xueyuan Street}, \city{Hangzhou}, \postcode{310018}, \state{Zhejiang Province}, \country{China}}}

\affil*[2]{\orgdiv{College of Information Engineering}, \orgname{China Jiliang University}, \orgaddress{\street{No. 258 Xueyuan Street}, \city{Hangzhou}, \postcode{310018}, \state{Zhejiang Province}, \country{China}}}

\affil[3]{\orgdiv{College of Economics and Management}, \orgname{China Jiliang University}, \orgaddress{\street{No. 258 Xueyuan Street}, \city{Hangzhou}, \postcode{310018}, \state{Zhejiang Province}, \country{China}}}


\abstract{External pressure significantly influences microcirculatory capillary blood flow, yet current studies lack quantitative modeling. This work proposes a nonlinear segmented coupling model between external pressure and capillary flow, incorporating vascular elasticity and collapse effects. The pressure–flow response is divided into three phases: elastic compression under low pressure ($<$30\text{mmHg} ), elliptical collapse in the transition zone ( 30$-$40\text{mmHg} ), and closure-induced attenuation under high pressure ($\geq$40\text{mmHg}), with explicit expressions derived for each. Parameter sensitivity analysis and comparison with literature demonstrate the model’s capability in capturing key determinants. The proposed framework supports dose–response assessment and individualized parameter tuning in pressure-based therapies such as tourniquets and compression garments.}




\maketitle

\section{Introduction}\label{sec1}

The microcirculation network consists of arterioles, capillaries, and venules. Among them, capillaries, with an average diameter of approximately 6–9 \textmu m, serve as critical conduits connecting arterioles and venules. Their reticular distribution provides highly efficient pathways for substance exchange between blood and tissues\citep{murrant2022}.Capillaries deliver oxygen and nutrients to cells and remove metabolic waste, playing a key role in maintaining internal homeostasis. As a result, capillary function has become a central focus of microcirculatory research\citep{roy2022}.

External pressure, such as pressure garments and microneedle interventions, is widely used in clinical applications. Studies have demonstrated that such pressure significantly influences capillary blood flow\citep{guven2020}\citep{debruler2019}. However, most existing studies focus primarily on experimental or simulation-based investigations and lack unified quantitative expressions, which makes it difficult to accurately assess therapeutic dosages\citep{farina2021}.

Vascular collapse under external pressure is a typical phenomenon of structural instability. The critical collapse pressure depends jointly on both internal blood pressure and external compression. Once this critical threshold is exceeded, vessel deformation leads to flow interruption\citep{sree2019}.In recent years, researchers have applied hemodynamic modeling to explore capillary flow characteristics and their regulation. For instance, Panula et al.\citep{panula2022} proposed a noninvasive method based on distal sampling to monitor blood flow dynamics, establishing theoretical links between structural and flow parameters. Tagliabue et al.\citep{tagliabue2023}used near-infrared scattering to study cerebral microcirculatory perfusion pressure, revealing a strong coupling between external pressure and microvascular flow changes. Linder-Ganz and Gefen\citep{linder2007} showed via simulation that increasing interstitial pressure leads to a progressive decline in capillary perfusion.

Therefore, this study establishes a quantitative model describing the relationship between external pressure and capillary blood flow by incorporating vessel elasticity and collapse effects. The model captures the nonlinear coupling between pressure and flow using piecewise representations, validated through sensitivity analysis and trend comparisons. We argue that this work not only provides a potential explanation for the clinically observed ``pressure overload paradox,'' but also offers a predictive tool for dose–response assessment in pressure-based therapies such as tourniquet application and anti-edema compression. This framework further enables personalized parameter optimization, balancing tissue perfusion with the risk of mechanical injury.

\section{Background and Theory}\label{sec2}

\subsection{Microcirculatory Capillary Structure}

The microcirculation comprises arterioles, capillaries, and venules, with capillaries serving as the primary site for mass exchange. Their diameter (6--9\,\textmu m) closely matches that of red blood cells ($\sim$8\,\textmu m), and their wall, composed of a single endothelial cell layer on a basement membrane, is only 0.5--1\,\textmu m thick~\citep{song2024}. This structure facilitates efficient exchange of gases and solutes~\citep{tooke2019microcirculation}. 

Flow within capillaries is highly stable due to the low Reynolds number ($Re \sim 10^{-4}$), well below the laminar--turbulent transition threshold~\citep{fusi2021}. The capillary wall’s low elastic modulus ($10^3$--$10^5$\,Pa) allows significant deformation, aiding pressure buffering~\citep{huang2024,kassab2019}. For small strains ($<$15\%), linear elasticity provides a valid approximation~\citep{baskurt2024blood}.

Blood in capillaries displays shear-thinning behavior. At high shear rates ($>$100\,s$^{-1}$), red blood cells migrate axially, forming a cell-free layer and reducing apparent viscosity to 2.5--3.5\,mPa$\cdot$s~\citep{loiseau2019}. Given the low $Re$ and near-steady flow, Stokes flow assumptions are valid, forming the basis for hemodynamic modeling~\citep{wanless2020}.

\subsection{Critical‑Pressure Theory of Vascular Collapse}

Vascular collapse occurs when external pressure exceeds the vessel wall's mechanical capacity, leading to buckling and lumen deformation. This critical condition depends on geometric parameters ($R$, $h$) and mechanical properties ($E$, $\nu$)~\citep{wanless2020,sharzehee2019}.

The classical Starling-resistor model defines collapse at $P_{\text{ext}} \ge P_{\text{int}}$, using a piecewise function where flow obeys Poiseuille’s law before collapse:
\[
Q = \frac{\pi \, \Delta P \, R^{4}}{8 \eta L}
\]
and drops toward a limit afterward~\citep{tong2019}. However, this neglects wall elasticity.

Including vessel elasticity via Laplace's law and linear elasticity gives:
\[
P_{\text{ext,crit}} = P_{\text{int}} + \frac{E h \,\varepsilon}{r\,(1 - \nu^{2})}
\]
where $\varepsilon$ is the circumferential strain~\citep{doh2021}. The internal pressure–radius relationship is:
\[
P_{\text{int}} = \frac{E h}{r_{0}}\left(\frac{r}{r_{0}} - 1\right)
\]
allowing analytical estimation of collapse thresholds.

Non-Newtonian properties of blood can be introduced by modeling viscosity $\eta$ as a function of shear rate~\citep{sloop2020}, though empirical data are required for calibration.

\subsection{Fundamentals of Hemodynamics}

Hemodynamics governs the interplay between blood flow and vascular structure. Poiseuille’s law describes steady laminar flow in rigid cylindrical tubes:
\[
Q = \frac{\pi \, \Delta P \, R^4}{8 \eta L}
\]
where $Q$ is flow rate, $\Delta P$ the pressure gradient, $R$ the radius, $\eta$ the viscosity, and $L$ the vessel length~\citep{doh2021}. However, microvessels can deform significantly under external pressure. When wall stress exceeds a critical threshold, the circular lumen may flatten into an ellipse or collapse entirely~\citep{thiriet2019,seddighi2021}, a process modeled via elastic shell buckling and Fung’s soft tissue mechanics~\citep{fung2013biomechanics}.

To account for deformation, we introduce an elliptical correction factor $\beta$, which adjusts hydraulic resistance for elliptical cross-sections~\citep{garg2024}. Physically, $\beta$ quantifies the relative increase in resistance compared to an undeformed circular vessel, enabling more accurate flow prediction under structural distortion.

\subsection{Physiological Interpretation of the Multi-Phase Pressure–Flow Relationship}

Mechanical modulation of microcirculation, such as by compression garments or microneedle stimulation, exhibits a characteristic multi-phase pressure–flow response, driven by dynamic vascular geometry changes.

At pressures below 30\,mmHg, vessels retain a nearly cylindrical shape due to endothelial regulation. Flow follows a linear pressure–flow relationship, consistent with Poiseuille’s law, governed by parameters such as vessel elasticity and blood viscosity.

As pressure exceeds $\sim$30\,mmHg, vessels deform elliptically~\citep{seddighi2021}. This transition increases hydraulic resistance and induces a nonlinear flow decline. The correction factor $\beta$, as a function of the major-to-minor axis ratio, captures the effect of geometric distortion on flow impedance~\citep{garg2024}. Red blood cells also face higher deformation and energy loss in narrowed regions.

Beyond $\sim$40\,mmHg, vessels enter a collapse-dominant stage. We introduce a pressure-dependent attenuation coefficient $\alpha$ to characterize the sharp flow drop due to lumen closure. Integrating these three stages, a segmented model is formulated to reflect morphological evolution, nonlinear elasticity, and flow coupling—detailed in the ``Model Development'' section.

\section{Model Development}\label{sec3}
Based on the theoretical background described above, we divide the relationship between capillary blood flow \( Q(q) \) and external pressure \( q \) into three stages: low-pressure region, transition region, and high-pressure region. Corresponding flow models are established for each stage. The derivation is as follows:

\subsection{Fundamental Theory and Modeling Assumptions}\label{subsec2}
Under steady-state laminar flow conditions, the blood flow inside capillaries can be approximated by Poiseuille:
\[
Q = \frac{\pi \Delta P R^4}{8 \eta L}
\]
where \( \Delta P \) is the pressure difference between the inside and outside of the vessel, \( R \) is the lumen radius, \( \eta \) is the apparent viscosity of blood, and \( L \) is the vessel length. Since this study focuses on the effect of external pressure on structural deformation, the internal pressure \( P_{\text{int}} \) is assumed to be approximately constant, and the external pressure \( q = P_{\text{ext}} \) is treated as the sole variable.

The initial capillary radius is defined as \( r_0 \), and the reference flow \( Q_0 \) is defined as the flow rate in the absence of any pressure disturbance:
\[
Q_0 = \frac{\pi r_0^4}{8 \eta L} \cdot \Delta P
\]

Based on this, the variation in flow is entirely attributed to the deformation of the effective radius \( R(q) \), and thus:
\[
Q(q) = Q_0 \left( \frac{R(q)}{r_0} \right)^4
\]

\subsection{Low-Pressure Region (\(q < 30\, \text{mmHg}\)): Linear Elastic Compression Phase}
In this phase, the capillary wall is only mildly compressed, exhibiting linear elastic compression behavior. The relative radial compression ratio can be expressed as:
\[
\frac{\Delta r}{r_0} = \frac{q r_0}{E h}
\]
Thus, the deformed radius becomes:
\[
R(q) = r_0 \left(1 - \frac{q r_0}{E h} \right)
\]

Substituting into the Poiseuille equation yields:
\[
Q(q) = Q_0 \left(1 - \frac{q r_0}{E h} \right)^4
\]

This expression reflects a compliance-dominated nonlinear attenuation effect under low external pressure.

Considering the active regulatory role of endothelial cells, the flow expression is modified when $q < q_{\text{yield}}$ as follows:
\[
Q(q) = Q_0 \left(1 - \frac{q}{q_{\text{yield}}} \right)^4
\]
Detailed rationale is provided in the \textit{Discussion} section.

\subsection{Transition Region (\(30 \leq q < 40\, \text{mmHg}\)): Elliptical Collapse Correction Phase}

As the external pressure further increases beyond a certain critical value (e.g., \(q > 30\, \text{mmHg}\)), the vascular cross-section begins to collapse from a circular to an elliptical shape. In this stage, flow resistance is no longer governed solely by radius reduction; instead, an elliptical correction factor \(\beta(q)\) must be introduced to reflect the impact of cross-sectional deformation on the equivalent hydraulic radius:
\[
Q(q) = Q_0 \left( 1 - \frac{q}{K_{\text{collapse}}} \right)^4 \cdot \beta(q)
\]

Here, $K_{\text{collapse}}$ denotes the critical stiffness parameter associated with vessel collapse, and $\beta(q)$ is a function of the aspect ratio $a/b$, representing the hydraulic attenuation due to elliptical cross-sectional deformation. It is defined as:

\[
\beta(q) = \frac{(a/b)^2}{1 + (a/b)^2}
\]

The geometric relationship is derived based on Han’s theory of elastic shell buckling\citep{han2004}, which enables the model to more accurately characterize the progressive deformation of the vascular cross-section from a circular to an elliptical shape during collapse.

\subsection{High-Pressure Region (\(q \geq 40\, \text{mmHg}\)): Collapse and Decay Phase}

When the external pressure exceeds the collapse threshold, the vascular cross-section tends toward closure, and the flow rate drops sharply to a minimal value. This phenomenon can be approximately modeled as an exponential decay process:
\begin{equation}
    Q(q) = Q_0 \cdot e^{-\alpha (q - 40)}
\end{equation}
where $\alpha$ is the decay coefficient, which should be determined based on experimental or physiological data. It typically reflects dependencies on tissue type, shear rate sensitivity, and other factors.

\subsection{Model Summary: Piecewise Flow–Pressure Function}
In summary, the complete model describing the relationship between capillary flow rate and external pressure can be expressed as:
\begin{equation}
Q(q) = 
\begin{cases} 
Q_0 \left(1 - \dfrac{q r_0}{E h} \right)^4, & q < 30\ \text{mmHg} \\
Q_0 \left(1 - \dfrac{q}{K_{\text{collapse}}} \right)^4 \cdot \beta(q), & 30 \leq q < 40\ \text{mmHg} \\
Q_0 \cdot e^{-\alpha (q - 40)}, & q \geq 40\ \text{mmHg}
\end{cases}
\end{equation}
The simulated flow-pressure curve is shown in Figure~1, capturing the distinct characteristics of each phase.
\begin{figure}[htbp]
    \centering
    \includegraphics[width=0.8\textwidth]{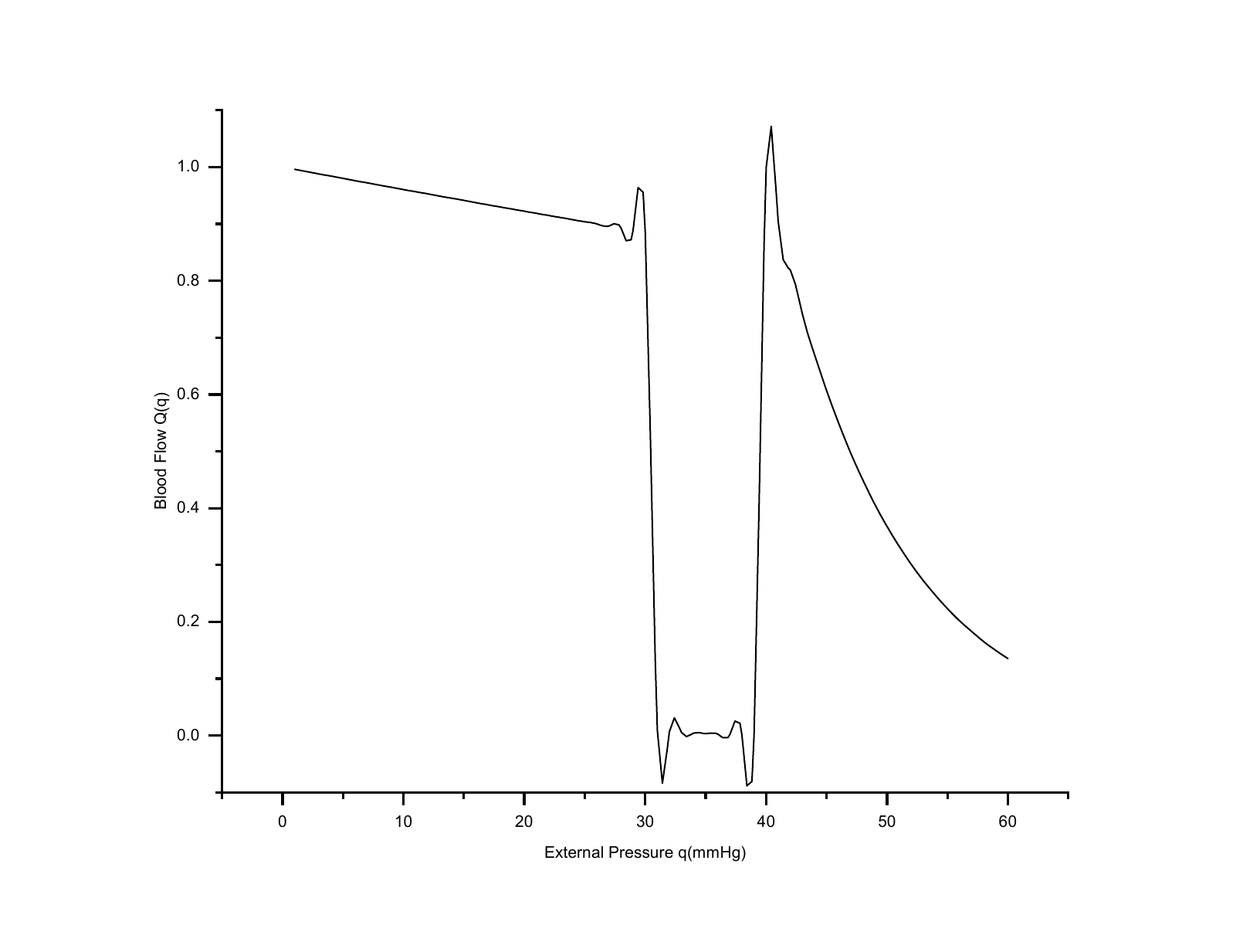}
    \caption{External Pressure-Blood Flow relationship diagram}
    \label{flow-curve}
\end{figure}

\section{Parameter Estimation and Sensitivity Analysis}\label{sec8}
\begin{figure}[htbp]
    \centering
    \begin{subfigure}[b]{0.45\textwidth}
        \includegraphics[width=\textwidth]{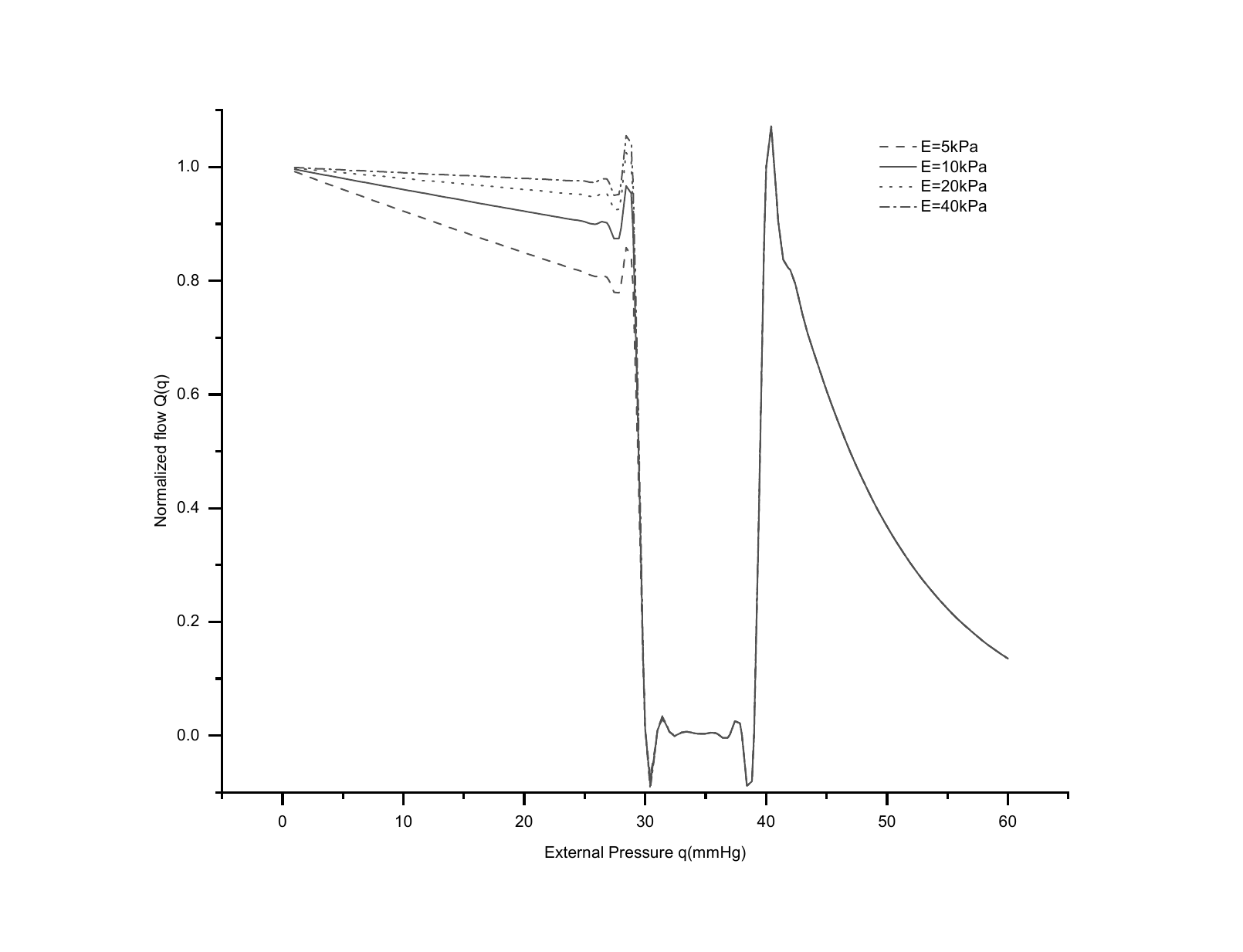}
        \caption{Sensitivity of Flow Q(q) to Elastic Modulus E}
        \label{fig:1a}
    \end{subfigure}
    \hfill
    \begin{subfigure}[b]{0.45\textwidth}
        \includegraphics[width=\textwidth]{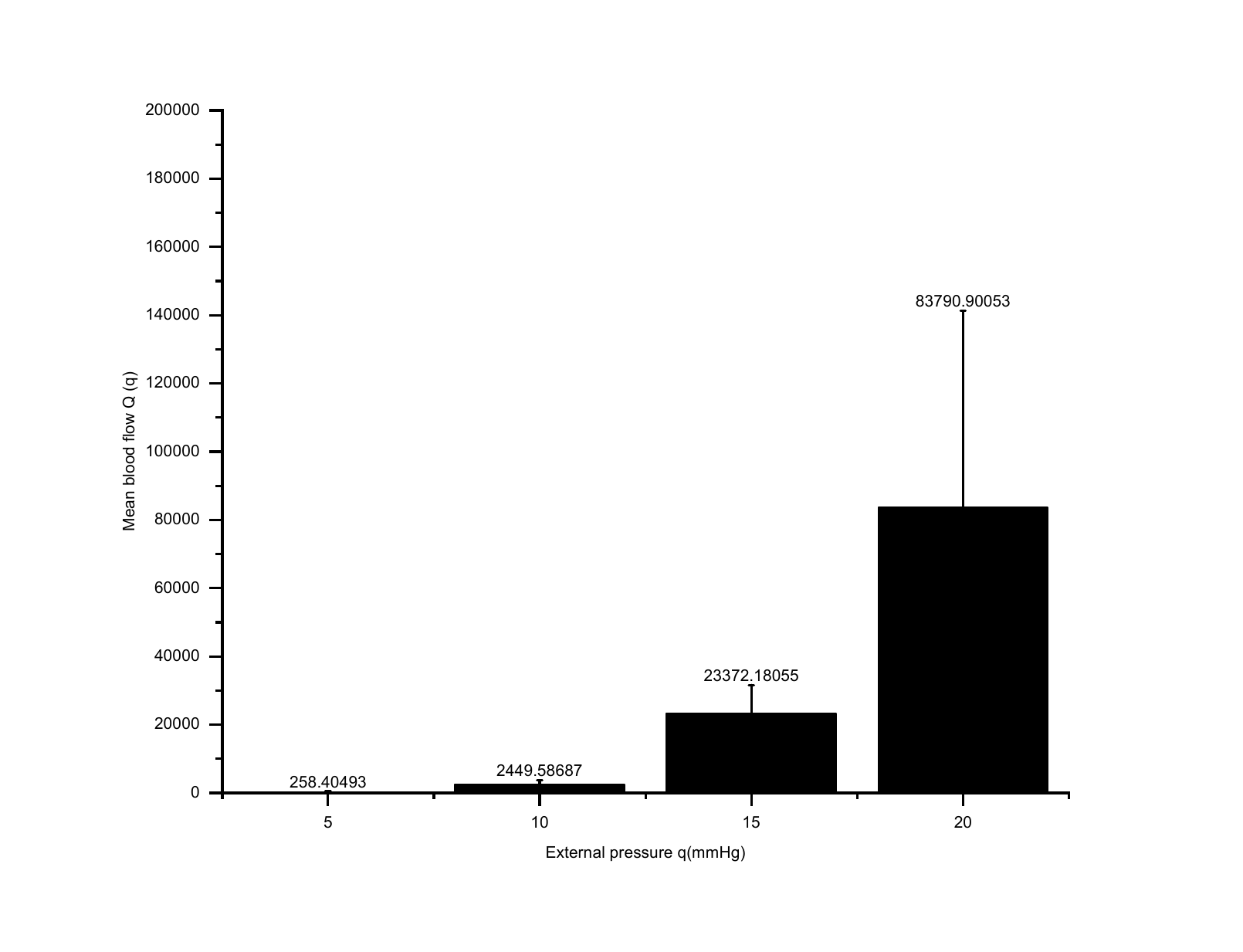}
        \caption{Error bar analysis of vascular elasticity parameter $E$ in the low-pressure region}
        \label{fig:1d}
    \end{subfigure}
    
    \vspace{0.5cm} 
    
    \begin{subfigure}[b]{0.45\textwidth}
        \includegraphics[width=\textwidth]{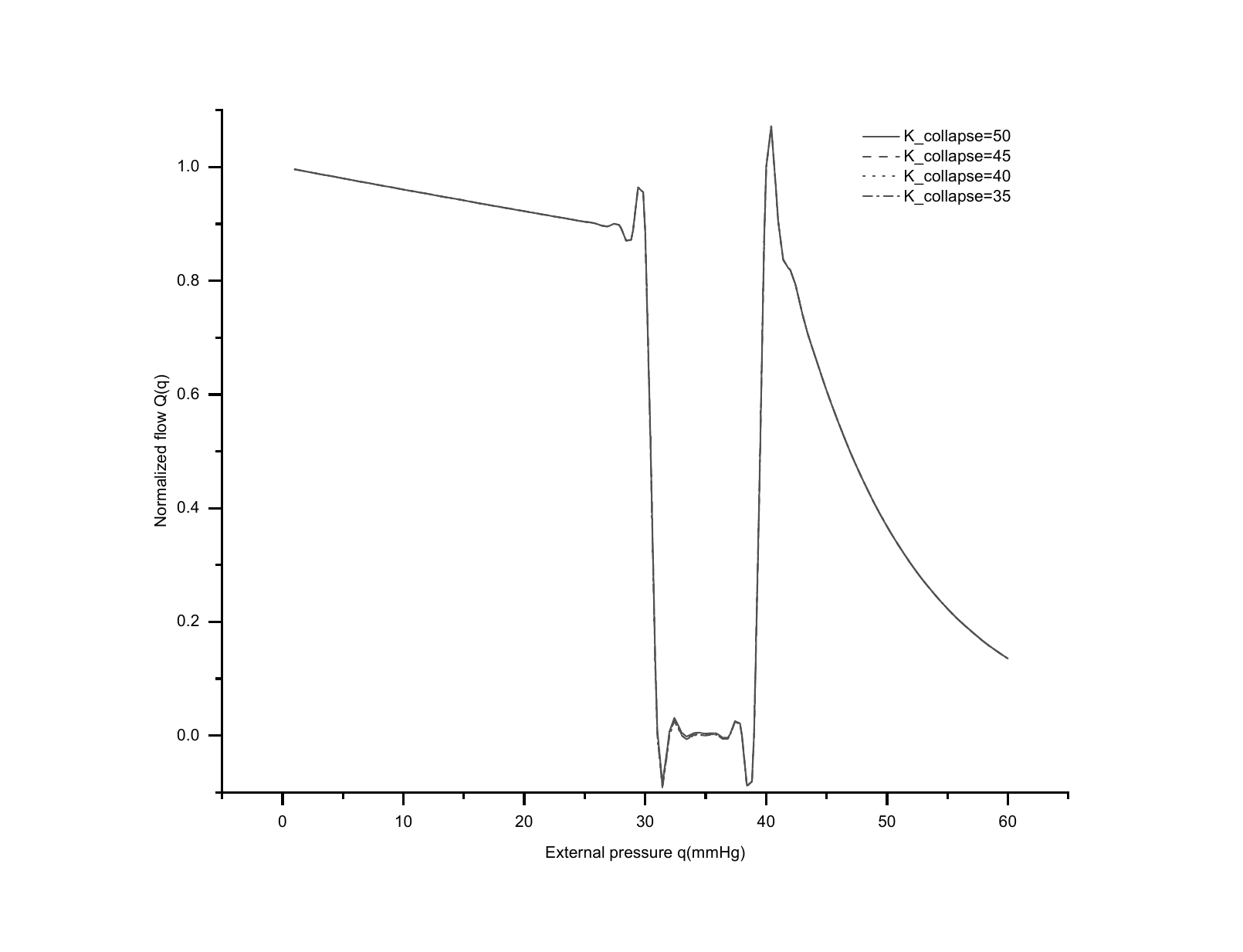}
        \caption{Sensitivity of Flow Q(q) to collapse parameter K}
        \label{fig:1b}
    \end{subfigure}
    \hfill
    \begin{subfigure}[b]{0.45\textwidth}
        \includegraphics[width=\textwidth]{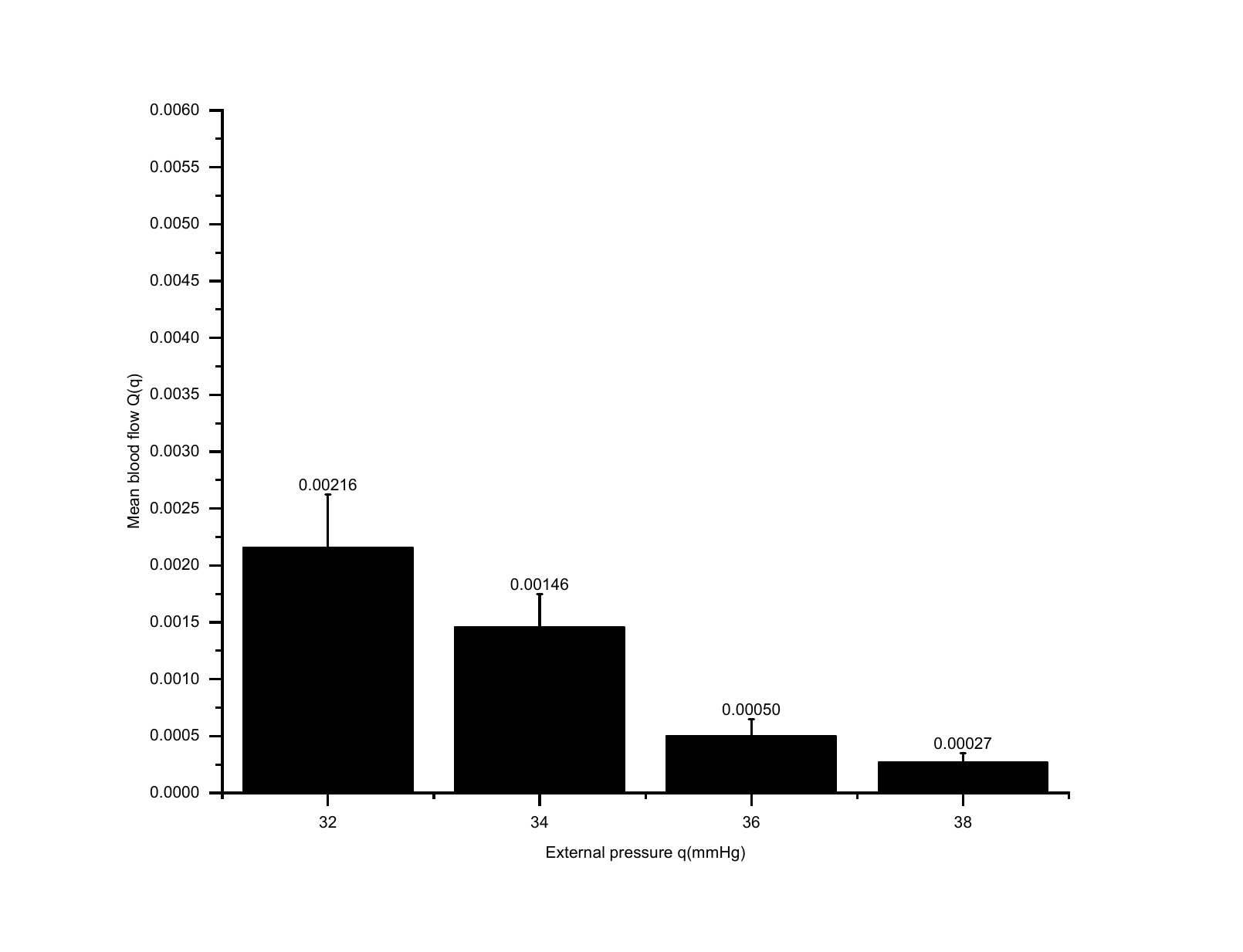}
        \caption{Error bar analysis of vascular collapse parameter $K$ in the transition region}
        \label{fig:1e}
    \end{subfigure}
    
    \vspace{0.5cm} 
    
    \begin{subfigure}[b]{0.45\textwidth}
        \includegraphics[width=\textwidth]{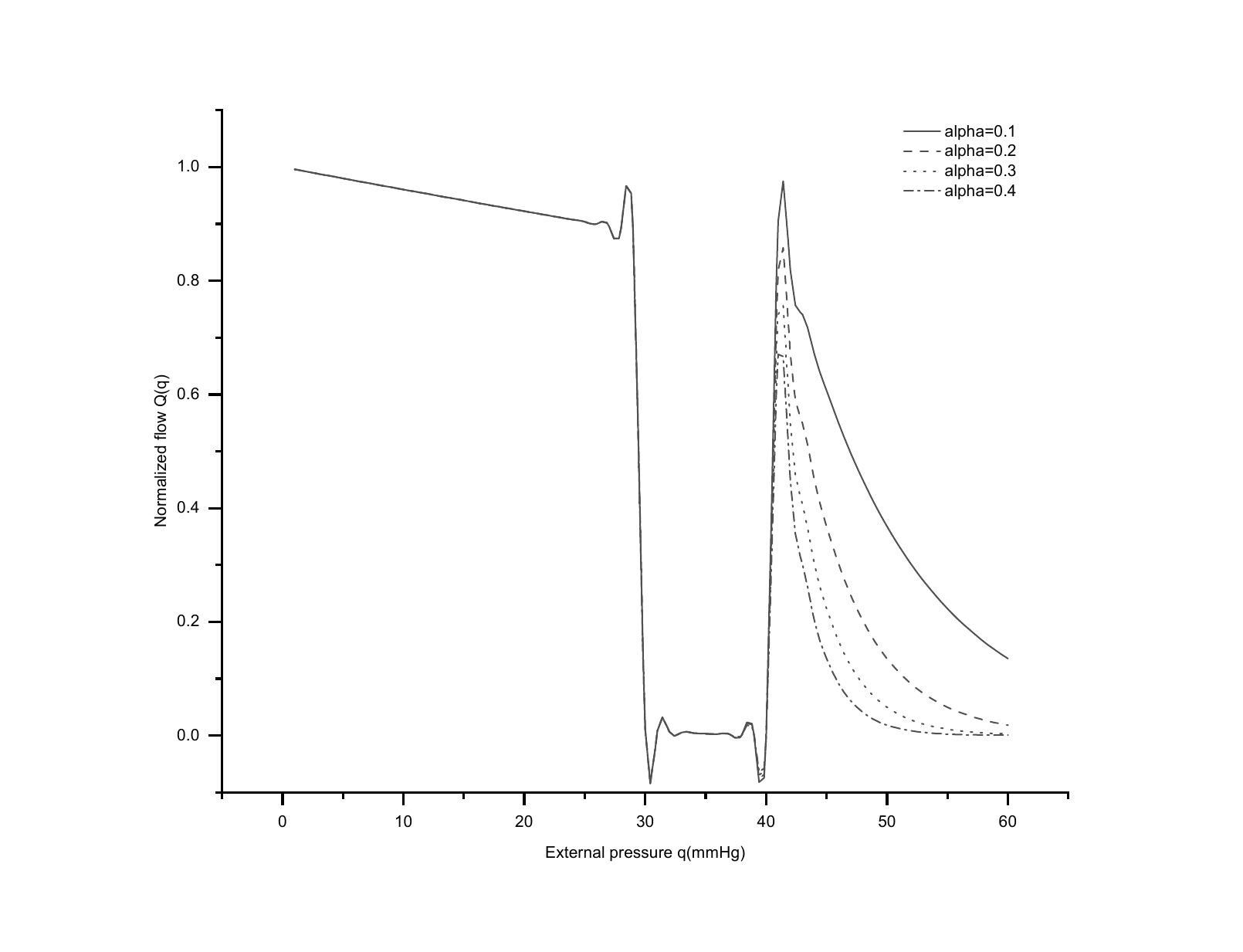}
        \caption{Sensitivity of Flow Q(q) to decay coefficient $\alpha$}
        \label{fig:1c}
    \end{subfigure}
    \hfill
    \begin{subfigure}[b]{0.45\textwidth}
        \includegraphics[width=\textwidth]{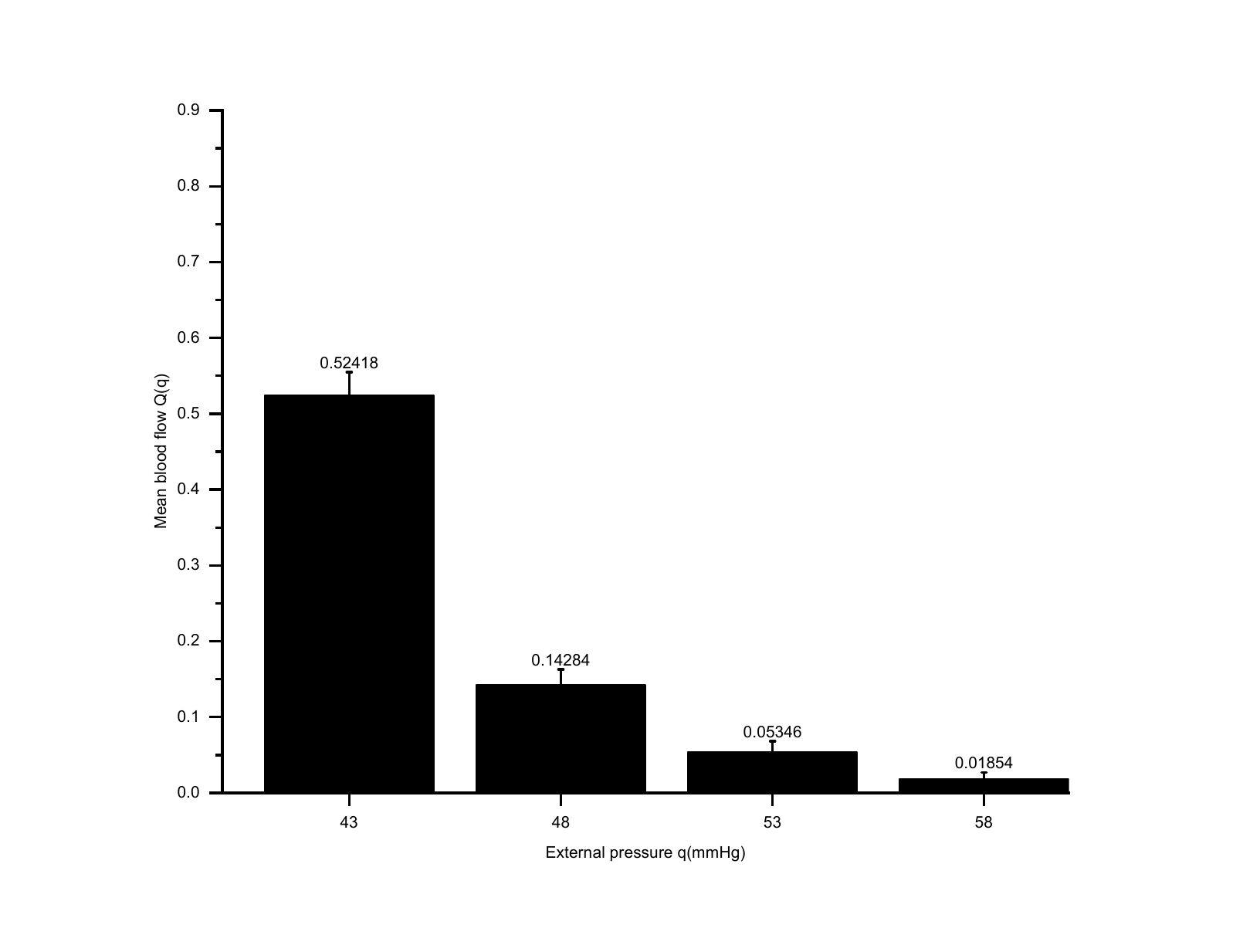}
        \caption{Error bar analysis of vascular decay coefficient $\alpha$ in the high-pressure region}
        \label{fig:1f}
    \end{subfigure}
    \caption{Sensitivity analysis of flow $Q(q)$ to key parameters. (a,c,e): effects of elastic modulus $E$, collapse parameter $K$, and decay coefficient $\alpha$, respectively. (b,d,f): error bar analysis of vascular parameters across different pressure regions—(b) $E$ in the low-pressure region, (d) $K$ in the transition-pressure region, and (f) $\alpha$ in the high-pressure region. Error bars indicate standard error of the mean (SEM), $n=20$; simulated data were used to illustrate model prediction variability.}

    \label{fig:combined}
\end{figure}

Under the default parameter conditions (Table~1), sensitivity analysis reveals that the hemodynamic response to external pressure $q$ is significantly influenced by the synergistic interaction of the elastic modulus $E$, collapse coefficient $k$, and decay coefficient $\alpha$. 

\begin{table}[h]
\centering
\caption{Experimental data source table}
\label{tab:experimental-data}
\begin{tabular}{lccc}
\toprule
Parameter & Value & Unit & Source \\
\midrule
$E$ & 10 & kPa & Literature reference$^{a}$ \\
$h$ & 0.001 & mm & Literature reference$^{b}$ \\
$K$ & 50 & mmHg & assumption \\
$\beta$ & 0.4 & none & fitted value \\
$r_0$ & 0.01 & mm & Experimental anatomical mean value \\
$\alpha$ & 0.1 & none & assumption \\
\bottomrule
\end{tabular}
\begin{tablenotes}
\small
\item[a] Reference \citep{abdullah2022}data
\item[b] Data source:\citep{braverman2000}
\end{tablenotes}
\end{table}

As shown in Figure~a, within the low-pressure range ($0 < q < 30$\,mmHg), the flow rate $Q(q)$ is positively correlated with $E$: when $E = 5$\,kPa, the vessel exhibits high compliance, resulting in significant geometric deformation under pressure and a rapid early decline in flow. In contrast, at $E = 40$\,kPa, the vessel exhibits greater resistance to compression, thereby maintaining better perfusion. This observation suggests that in individuals with reduced microvascular compliance (e.g., elderly or diabetic patients), blood flow may decline earlier even under pressures below 30\,mmHg. Conversely, enhancing vascular tone to increase tissue elasticity may delay the onset of flow impairment, highlighting potential clinical value for intervention.

Once external pressure exceeds 30\,mmHg, vascular impedance and geometric deformation begin to dominate the hemodynamic response, and the flow rate becomes markedly less sensitive to changes in $E$. This supports the notion that elastic response primarily governs the model behavior in the low-pressure regime, consistent with the three-phase coupling framework.

However, analysis of the mean and standard error of capillary flow across different pressure groups (Figure~b) shows a marked increase in data variability with increasing pressure (e.g., at $q = 15$\,mmHg and $q = 20$\,mmHg). This indicates potential discrepancies between model predictions and actual physiological values under certain conditions. The pronounced differences in error bar lengths further suggest limitations in the model's ability to accurately describe hemodynamic behavior under higher pressures within this range.

In the transitional pressure range (\(30 \leq q < 40\, \text{mmHg}\)), the collapse coefficient $K$ governs the dynamic characteristics of flow reduction (Figure~c). When $K = 35$\,mmHg, the vessel exhibits a relatively low stiffness threshold, resulting in a near-abrupt drop in flow just above 30\,mmHg. In contrast, when $K = 50$\,mmHg, the decline in flow becomes more gradual. This suggests that $K$ can be interpreted as the critical stiffness reflecting capillary resistance to external pressure: under conditions of vascular elastic degradation or insufficient intravascular pressure, a smaller $K$ value indicates earlier onset of nonlinear collapse in flow. Conversely, vascular stiffening or elevated internal pressure is represented by a larger $K$, conferring greater resistance to pressure-induced deformation.

According to the analysis of error bar lengths and distributions (Figure~d), the model predictions of capillary blood flow within this transitional pressure range show high credibility. Under four different pressure conditions, the flow rate $Q(q)$ decreased in a stepwise pattern (from 0.00216 to 0.00027), and the narrow range of error bars at each data point indicates low dispersion among measurements. This implies good repeatability of simulation outputs and limited influence from within-group variability.

In the high-pressure range (\(q \geq 40\, \text{mmHg}\)), the decay coefficient $\alpha$ modulates the rate of flow reduction (Figure~e). As $\alpha$ increases from 0.1 to 0.4, the slope of the flow curve becomes significantly steeper: at $q = 50$\,mmHg, flow declines sharply from approximately 0.36 to less than 0.02. This indicates that higher $\alpha$ values correspond to weaker residual perfusion capacity following vascular closure, while smaller $\alpha$ values suggest that collateral circulation can sustain partial flow.

The error bar distribution (Figure~f) further supports the reliability of the model in the high-pressure regime. Across four pressure conditions, the mean flow values show a clear decreasing trend with increasing pressure (from 0.5 to 0.02), and the error bar range narrows progressively with pressure. Together, the sensitivity analysis and error bar evaluations suggest that $K$ and $\alpha$ serve as critical parameters for individualized modeling. By aligning these parameters with specific pathological conditions, the model provides a theoretical foundation for the precise assessment of clinical pressure intervention strategies and safety thresholds.

\begin{figure}[htbp]
    \centering
    \includegraphics[width=0.8\textwidth]{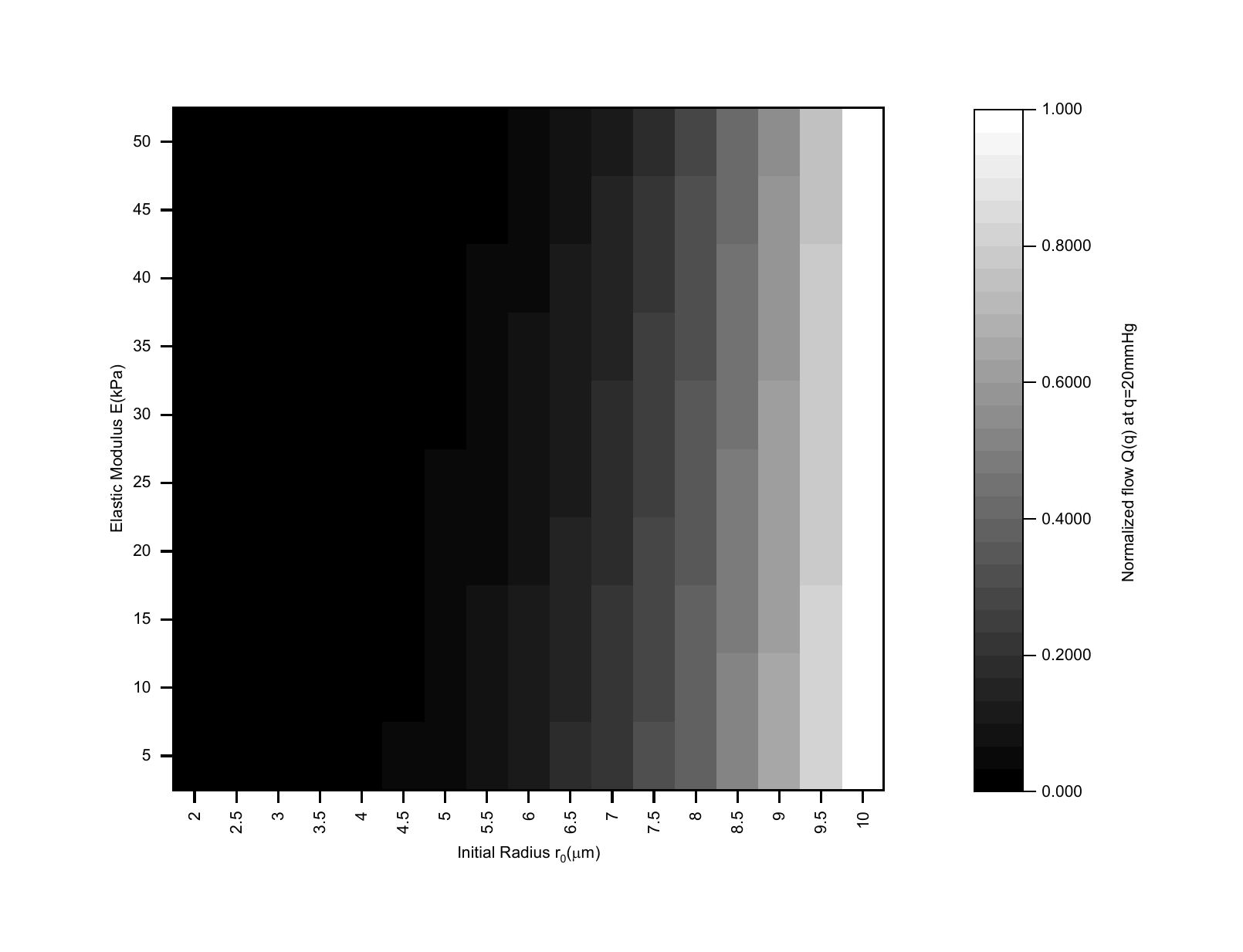}
    \caption{Sensitivity Map:r0 vs E(at q= 20 mmHg)}
    \label{Sensitivity Map}
\end{figure}

As shown in Figure~3, the effects of initial vessel radius $r_0$ and elastic modulus $E$ on flow rate within the low-pressure range exhibit a distinct nonlinear response. In the lower-left region of the contour map—corresponding to small values of both $r_0$ and $E$—a large area appears nearly black, indicating that the normalized flow approaches zero. This suggests a high susceptibility to functional occlusion under such parameter combinations. In contrast, in the upper-right region (i.e., larger $r_0$ and higher $E$), the flow remains elevated, approaching a normalized value of 1, implying that sufficient diameter and stiffness help sustain unobstructed blood flow.

Notably, the transition between high-flow and low-flow regions is not linear but forms a distinct nonlinear gradient boundary, highlighting the complex coupling between the two parameters. The asymmetry of this transition zone shows some deviation from actual physiological conditions. Specifically, in parameter combinations involving small, compliant vessels, the model predicts an excessively rapid flow reduction, which does not fully align with experimental observations indicating that capillaries can still maintain a degree of perfusion under normal physiological conditions. The underlying causes of this discrepancy will be further explored in the \textit{Discussion} section.

\begin{figure}[htbp]
    \centering
    \includegraphics[width=0.8\textwidth]{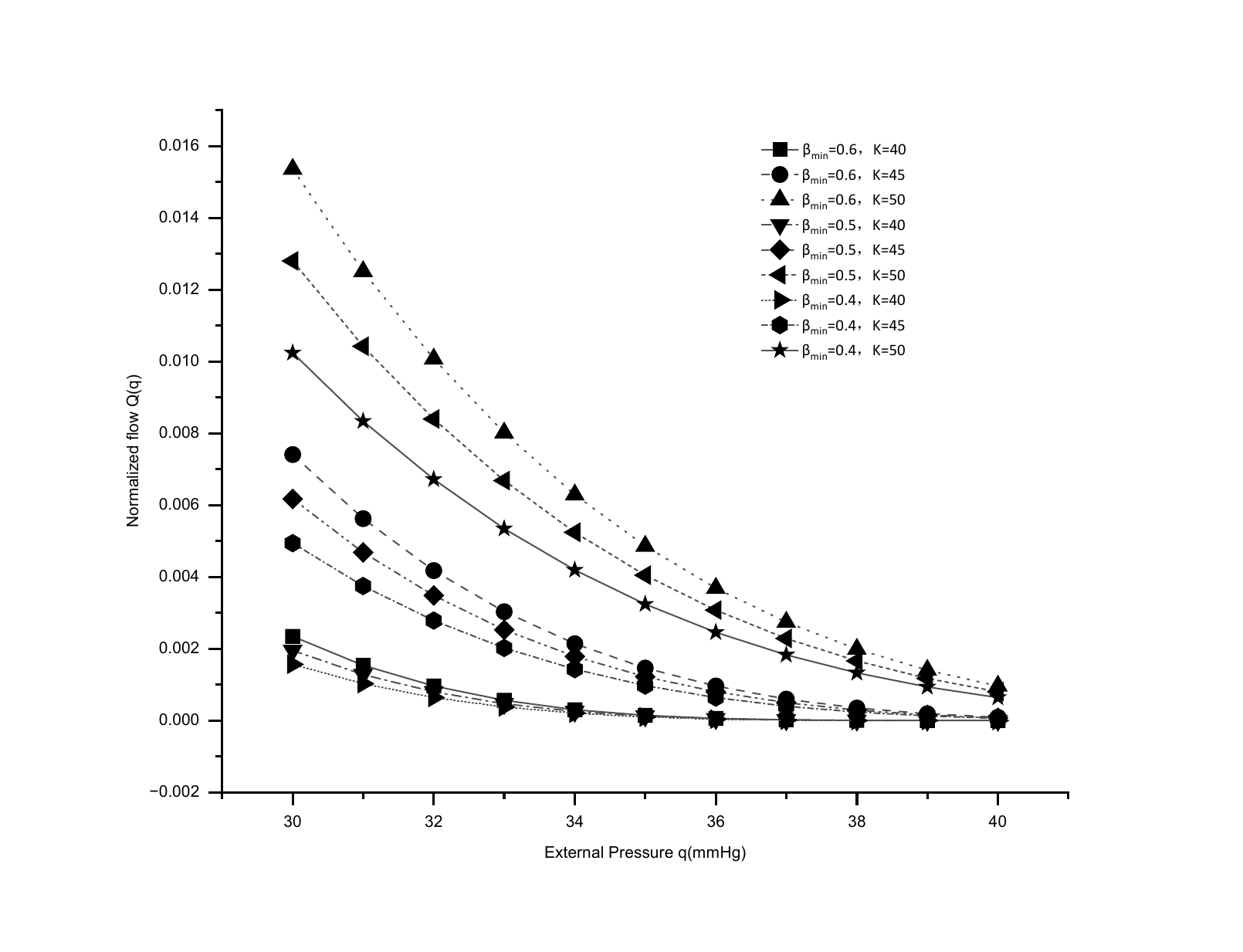}
    \caption{Transition phase: $B(q)$–$K_{\text{Collapse}}$ interaction}
    \label{B(q)-K_Collapse Interaction}
\end{figure}

Within the external pressure range of $30 \leq q < 40\, \text{mmHg}$, the pressure–flow relationship exhibits pronounced nonlinear characteristics (Figure~4). This range represents a critical transition phase during which vessels evolve from linear deformation into pronounced collapse. In our model, both the elliptical correction factor $\beta$ and the collapse stiffness coefficient $K$ jointly modulate flow behavior in this zone.

When the collapse coefficient $K$ is held constant, increasing the elliptical correction factor $\beta$ from 0.4 to 0.6 results in a more rapid flow decline. This phenomenon suggests that a higher $\beta$ corresponds to enhanced cross-sectional ellipticity post-collapse (e.g., due to fibrosis or edema), which substantially reduces the effective lumen area and increases hydraulic resistance. Consequently, even when the vessel is not fully occluded, perfusion capacity is already severely compromised. Conversely, for a fixed $\beta$, smaller values of $K$ lead to more gradual pressure–flow curves. This indicates that vessels enter the collapse state earlier, but the collapse progresses more slowly. Such gradual deformation may reflect underlying physiological mechanisms such as collagen remodeling or structural damping that delay critical flow reduction.

Importantly, the interaction between $\beta$ and $K$ is not a simple linear superposition. Under pathological simulation conditions (e.g., $\beta = 0.6$, $K = 40$), the most severe flow decay is observed, indicating that geometric deterioration ($\beta \uparrow$) and reduced mechanical resistance ($K \downarrow$) have a synergistic detrimental effect. Nevertheless, the model also reveals a potential interventional pathway: increasing the value of $K$ may partially offset the adverse impact of elevated $\beta$. For instance, when $K$ increases from 40\,mmHg to 45\,mmHg, the pressure–flow curve for $\beta = 0.6$ approximates that of the reference case with $\beta = 0.4$ and $K = 45\,\text{mmHg}$ around $q \approx 38\,\text{mmHg}$. Although this result emerges from modeling assumptions, it implies that in the context of vascular degeneration, mechanical reinforcement strategies—such as increasing wall tension or structural support—may delay the collapse process and improve perfusion outcomes.

Clinically, these findings may provide theoretical support for interventions involving medical compression stockings~\citep{rastel2019}, external pressure devices, and other biomechanical aids. They suggest that even when geometric deterioration cannot be reversed, hemodynamic performance may still be optimized through mechanical regulation.

\section{Model Validation via Literature Comparison}\label{sec10}

Literature validation indicates that the relationship between external pressure applied to capillaries and blood flow can be reasonably explained by a three-phase model. In the low-pressure region (q $<$30 mmHg), the linear elastic deformation caused by mild vessel compression is consistent with the positive correlation between vessel diameter and blood flow proposed in mathematical models \citep{abdullah2022}. The physiological phenomenon where moderate negative pressure ($-$125 mmHg) optimizes perfusion further supports this stage \citep{negosanti2012}, though the correspondence between negative pressures in the literature and the positive pressure range used in this model requires further validation.

In the transition region (\(30 \leq q < 40\, \text{mmHg}\)), the interplay between vessel collapse and decreased blood flow is supported by previous findings \citep{langemo2008,sree2019phd}. External pressure has been shown to directly induce vascular collapse and ischemic injury, such as heel pressure ulcers \citep{langemo2008}, while multiscale modeling highlights pressure-driven vessel collapse as a key mechanism of pressure ulcer formation \citep{sree2019phd}.

In the high-pressure region (\(q \geq 40\, \text{mmHg}\)), the exponential decay characteristic is consistent with pathological processes where hypertension induces capillary collapse and structural vascular damage \citep{stompor2020}. Sustained high pressure may initiate a vicious cycle of perfusion block and hypoxia \citep{sree2019phd}, and the particular vulnerability of capillaries to pressure due to their structural fragility provides a physiological basis for the observed rapid decline in blood flow \citep{abdullah2022}.

It should be noted that the model’s critical thresholds (30 mmHg and 40 mmHg) currently lack direct clinical validation. These values are primarily based on experimental findings related to pressure-induced vascular collapse\citep{langemo2008,sree2019phd}. Future studies could verify these thresholds using microcirculatory imaging techniques such as optical coherence tomography (OCT) or laser speckle contrast imaging\citep{heeman2019}.

Furthermore, the physiological interpretation of the decay coefficient $\alpha$ refers to the link between hypertension and vascular injury \citep{stompor2020}. Its clinical relevance for pressure management in microcirculatory disorders—such as the use of intermittent pressure relief strategies—still requires validation through dynamic blood flow monitoring under different pressure conditions.

\section{Discussion}\label{sec12}

This study introduces a phase-based model that captures the nonlinear modulation of capillary blood flow by external pressure, divided into three distinct regimes: elastic compression ($q < 30$\,mmHg), transitional collapse (30--40\,mmHg), and high-pressure occlusion (\(q \geq 40\, \text{mmHg}\)). The model links structural deformation with hemodynamic response and offers a theoretical framework for optimizing pressure-based interventions.

A critical reduction in flow is observed above 30\,mmHg, suggesting a physiological threshold potentially tied to the elastic limit of the vessel wall. This may represent a self-protective mechanism to prevent rupture or endothelial injury. Moreover, flow decline beyond this point may promote the accumulation of metabolites (e.g., adenosine, lactate), triggering vasodilatory feedback to restore perfusion. Clinically, maintaining external pressure below this threshold is crucial in applications such as compression garments, where prolonged occlusion may lead to ischemic damage, while exceeding it briefly may support hemostasis.

In the transitional range (30--40\,mmHg), the nonlinear flow response supports the need for dynamic modulation in negative-pressure therapies. Controlled pressure fluctuations may enhance drainage while preventing microvascular collapse. However, individual variability in vascular stiffness, particularly in aging or diabetic populations, suggests that personalized pressure strategies are needed. Future studies could incorporate local stiffness measurements (e.g., via atomic force microscopy) to tailor interventions~\citep{orusbiev2023}.

At high pressures (\(q \geq 40\, \text{mmHg}\)), flow decays exponentially, indicating critical vascular collapse. This behavior aligns with clinical observations such as the ``pressure overload paradox,'' where excessive compression reduces perfusion~\citep{negosanti2012}. In diabetic foot management, repeated exceedance of this threshold—e.g., from poorly designed footwear—may accelerate ulceration or fibrosis~\citep{langemo2008}.

Despite capturing key physiological transitions, the model has limitations. In the low-pressure regime, certain parameter combinations (e.g., large $r_0$, low $E$) may cause the flow term $\left(1 - \frac{q r_0}{E h} \right)^4$ to approach zero prematurely. While not reflective of actual perfusion cessation, this highlights the model’s sensitivity to parameter values and its simplification of vascular structure.

The current model assumes capillaries as ideal elastic tubes and omits key features such as yield stress, tension buffering, and endothelial regulation. To address this, the low-pressure equation can be modified to include a yield threshold:

\[
Q(q) = Q_0 \left(1 - \frac{q}{q_{\text{yield}}} \right)^4,
\]

where $q_{\text{yield}}$ (e.g., 20\,mmHg) represents the upper limit of active endothelial tension maintenance~\citep{peng2019}. For $q \geq q_{\text{yield}}$, the model reverts to passive elastic compression.

Additionally, blood was modeled as a Newtonian fluid, neglecting shear-thinning effects observed in vivo~\citep{sun2023}. The predicted flow drop near zero in the low-pressure range reflects this structural emphasis rather than computational error. Future work should incorporate non-Newtonian viscosity, such as the Carreau model, to improve physiological fidelity.

In the transitional (30--40\,mmHg) and high-pressure (40--60\,mmHg) ranges, the model demonstrates stable performance and physiological consistency. Although formal significance testing was not conducted, the limited overlap among error bars suggests a plausible correlation between pressure gradients and flow reduction. Notably, the error-to-mean ratio remains below 30\% across all pressure levels, indicating acceptable variability and lending indirect support to model robustness.

These results suggest the model is capable of reliably capturing capillary collapse dynamics in medium-to-high pressure contexts, consistent with prior reports. However, additional validation using larger datasets or formal statistical analysis is needed to improve generalizability.

In summary, while the model shows strong physiological relevance under elevated pressures, its high sensitivity in the low-pressure domain reflects a trade-off between structural simplicity and mechanistic detail. This feature also highlights opportunities for future refinement, such as integrating oxygen transport models or metabolic feedback mechanisms to expand its applicability.

\section{Conclusion}\label{sec13}

In this paper, we construct a mathematical model coupling capillary blood flow and pressure based on fluid dynamics principles, quantitatively characterizing the nonlinear relationship between capillary hemodynamic parameters and externally applied pressure. This model provides a theoretical framework for pressure regulation in the microcirculation. In line with the development trend of artificial intelligence, the model can be integrated with parameterized feature extraction and machine learning algorithms to build a lightweight blood flow prediction system. In clinical scenarios such as chronic wound care (e.g., burn treatment), this technology is expected to offer dynamic data support for optimizing pressure regulation strategies and provide new insights for the intelligent development of microcirculation assessment tools.

\backmatter

\bmhead{Acknowledgements}

This work was supported by the Innovation and Entrepreneurship Training Program for College Students in Zhejiang Province (New Talent Program for College Students) under Grant No. 2025R409B041.

\section*{Declarations}

\begin{itemize}
\item Funding: Our project is funded by the “Innovation and Entrepreneurship
Training Program for College Students in Zhejiang Province” (New
Talent Program for College Students), with the project approval number
2025R409B041.
\item Conflict of interest: The authors declare no competing interests.
\end{itemize}





\bibliography{main}


\begin{thebibliography}{37}
\ifx \bisbn   \undefined \def \bisbn  #1{ISBN #1}\fi
\ifx \binits  \undefined \def \binits#1{#1}\fi
\ifx \bauthor  \undefined \def \bauthor#1{#1}\fi
\ifx \batitle  \undefined \def \batitle#1{#1}\fi
\ifx \bjtitle  \undefined \def \bjtitle#1{#1}\fi
\ifx \bvolume  \undefined \def \bvolume#1{\textbf{#1}}\fi
\ifx \byear  \undefined \def \byear#1{#1}\fi
\ifx \bissue  \undefined \def \bissue#1{#1}\fi
\ifx \bfpage  \undefined \def \bfpage#1{#1}\fi
\ifx \blpage  \undefined \def \blpage #1{#1}\fi
\ifx \burl  \undefined \def \burl#1{\textsf{#1}}\fi
\ifx \doiurl  \undefined \def \doiurl#1{\url{https://doi.org/#1}}\fi
\ifx \betal  \undefined \def \betal{\textit{et al.}}\fi
\ifx \binstitute  \undefined \def \binstitute#1{#1}\fi
\ifx \binstitutionaled  \undefined \def \binstitutionaled#1{#1}\fi
\ifx \bctitle  \undefined \def \bctitle#1{#1}\fi
\ifx \beditor  \undefined \def \beditor#1{#1}\fi
\ifx \bpublisher  \undefined \def \bpublisher#1{#1}\fi
\ifx \bbtitle  \undefined \def \bbtitle#1{#1}\fi
\ifx \bedition  \undefined \def \bedition#1{#1}\fi
\ifx \bseriesno  \undefined \def \bseriesno#1{#1}\fi
\ifx \blocation  \undefined \def \blocation#1{#1}\fi
\ifx \bsertitle  \undefined \def \bsertitle#1{#1}\fi
\ifx \bsnm \undefined \def \bsnm#1{#1}\fi
\ifx \bsuffix \undefined \def \bsuffix#1{#1}\fi
\ifx \bparticle \undefined \def \bparticle#1{#1}\fi
\ifx \barticle \undefined \def \barticle#1{#1}\fi
\bibcommenthead
\ifx \bconfdate \undefined \def \bconfdate #1{#1}\fi
\ifx \botherref \undefined \def \botherref #1{#1}\fi
\ifx \url \undefined \def \url#1{\textsf{#1}}\fi
\ifx \bchapter \undefined \def \bchapter#1{#1}\fi
\ifx \bbook \undefined \def \bbook#1{#1}\fi
\ifx \bcomment \undefined \def \bcomment#1{#1}\fi
\ifx \oauthor \undefined \def \oauthor#1{#1}\fi
\ifx \citeauthoryear \undefined \def \citeauthoryear#1{#1}\fi
\ifx \endbibitem  \undefined \def \endbibitem {}\fi
\ifx \bconflocation  \undefined \def \bconflocation#1{#1}\fi
\ifx \arxivurl  \undefined \def \arxivurl#1{\textsf{#1}}\fi
\csname PreBibitemsHook\endcsname

\bibitem[\protect\citeauthoryear{Abdullah}{2022}]{abdullah2022}
\begin{barticle}
\bauthor{\bsnm{Abdullah}, \binits{E.Y.}}:
\batitle{Study of pressure applied to blood vessels using a mathematical model}.
\bjtitle{International Journal of Nonlinear Analysis and Applications}
\bvolume{13}(\bissue{1}),
\bfpage{1341}--\blpage{1350}
(\byear{2022})
\end{barticle}
\endbibitem

\bibitem[\protect\citeauthoryear{Baskurt and Meiselman}{2024}]{baskurt2024blood}
\begin{bchapter}
\bauthor{\bsnm{Baskurt}, \binits{O.K.}},
\bauthor{\bsnm{Meiselman}, \binits{H.J.}}:
\bctitle{Blood rheology and hemodynamics}.
In: \bbtitle{Seminars in Thrombosis and Hemostasis},
vol. \bseriesno{50},
pp. \bfpage{902}--\blpage{915}
(\byear{2024}).
\bcomment{Thieme Medical Publishers, Inc.}
\end{bchapter}
\endbibitem

\bibitem[\protect\citeauthoryear{Braverman}{2000}]{braverman2000}
\begin{barticle}
\bauthor{\bsnm{Braverman}, \binits{I.M.}}:
\batitle{The cutaneous microcirculation}.
\bjtitle{Journal of Investigative Dermatology Symposium Proceedings}
\bvolume{5}(\bissue{1}),
\bfpage{3}--\blpage{9}
(\byear{2000})
\end{barticle}
\endbibitem

\bibitem[\protect\citeauthoryear{DeBruler et~al.}{2019}]{debruler2019}
\begin{barticle}
\bauthor{\bsnm{DeBruler}, \binits{D.M.}},
\bauthor{\bsnm{Baumann}, \binits{M.E.}},
\bauthor{\bsnm{Blackstone}, \binits{B.N.}},
\bauthor{\bsnm{Zbinden}, \binits{J.C.}},
\bauthor{\bsnm{McFarland}, \binits{K.L.}},
\bauthor{\bsnm{Bailey}, \binits{J.K.}},
\bauthor{\bsnm{Powell}, \binits{H.M.}}:
\batitle{Role of early application of pressure garments following burn injury and autografting}.
\bjtitle{Plastic and Reconstructive Surgery}
\bvolume{143}(\bissue{2}),
\bfpage{310}--\blpage{321}
(\byear{2019})
\end{barticle}
\endbibitem

\bibitem[\protect\citeauthoryear{Doh et~al.}{2021}]{doh2021}
\begin{barticle}
\bauthor{\bsnm{Doh}, \binits{G.}},
\bauthor{\bsnm{Kim}, \binits{B.}},
\bauthor{\bsnm{Lee}, \binits{D.}},
\bauthor{\bsnm{Yoon}, \binits{J.}},
\bauthor{\bsnm{Lim}, \binits{S.}},
\bauthor{\bsnm{Han}, \binits{Y.S.}},
\bauthor{\bsnm{Eo}, \binits{S.}}:
\batitle{Hemodynamic principles in free tissue transfer: Vascular changes at the anastomosis site}.
\bjtitle{Archives of Hand and Microsurgery}
\bvolume{26}(\bissue{4}),
\bfpage{285}--\blpage{292}
(\byear{2021})
\end{barticle}
\endbibitem

\bibitem[\protect\citeauthoryear{Dharmangadan~Sree}{2019}]{sree2019phd}
\begin{botherref}
\oauthor{\bsnm{Dharmangadan~Sree}, \binits{V.}}:
Multiscale and multiphysics modeling of pressure driven ischemia and ulcer formation in the skin.
PhD thesis,
Purdue University Graduate School
(2019)
\end{botherref}
\endbibitem

\bibitem[\protect\citeauthoryear{Farina et~al.}{2021}]{farina2021}
\begin{barticle}
\bauthor{\bsnm{Farina}, \binits{A.}},
\bauthor{\bsnm{Fasano}, \binits{A.}},
\bauthor{\bsnm{Rosso}, \binits{F.}}:
\batitle{Mathematical models for some aspects of blood microcirculation}.
\bjtitle{Symmetry}
\bvolume{13}(\bissue{6}),
\bfpage{1020}
(\byear{2021})
\end{barticle}
\endbibitem

\bibitem[\protect\citeauthoryear{Fusi et~al.}{2021}]{fusi2021}
\begin{barticle}
\bauthor{\bsnm{Fusi}, \binits{L.}},
\bauthor{\bsnm{Farina}, \binits{A.}},
\bauthor{\bsnm{Saccomandi}, \binits{G.}}:
\batitle{Linear stability analysis of the poiseuille flow of a stratified non-newtonian suspension: Application to microcirculation}.
\bjtitle{Journal of Non-Newtonian Fluid Mechanics}
\bvolume{287},
\bfpage{104464}
(\byear{2021})
\end{barticle}
\endbibitem

\bibitem[\protect\citeauthoryear{Fung}{2013}]{fung2013biomechanics}
\begin{bbook}
\bauthor{\bsnm{Fung}, \binits{Y.-c.}}:
\bbtitle{Biomechanics: Mechanical Properties of Living Tissues}.
\bpublisher{Springer},
\blocation{New York}
(\byear{2013})
\end{bbook}
\endbibitem

\bibitem[\protect\citeauthoryear{Guven et~al.}{2020}]{guven2020}
\begin{barticle}
\bauthor{\bsnm{Guven}, \binits{G.}},
\bauthor{\bsnm{Hilty}, \binits{M.P.}},
\bauthor{\bsnm{Ince}, \binits{C.}}:
\batitle{Microcirculation: physiology, pathophysiology, and clinical application}.
\bjtitle{Blood Purification}
\bvolume{49}(\bissue{1-2}),
\bfpage{143}--\blpage{150}
(\byear{2020})
\end{barticle}
\endbibitem

\bibitem[\protect\citeauthoryear{Garg et~al.}{2024}]{garg2024}
\begin{botherref}
\oauthor{\bsnm{Garg}, \binits{A.}},
\oauthor{\bsnm{Mishra}, \binits{H.}},
\oauthor{\bsnm{Pattanayek}, \binits{S.K.}}:
Optimal flow and scaling laws for power-law fluids in elliptical cross-section self-similar tree-like networks
(2024)
\end{botherref}
\endbibitem

\bibitem[\protect\citeauthoryear{Han et~al.}{2004}]{han2004}
\begin{barticle}
\bauthor{\bsnm{Han}, \binits{J.H.}},
\bauthor{\bsnm{Kardomateas}, \binits{G.A.}},
\bauthor{\bsnm{Simitses}, \binits{G.J.}}:
\batitle{Elasticity, shell theory and finite element results for the buckling of long sandwich cylindrical shells under external pressure}.
\bjtitle{Composites Part B: Engineering}
\bvolume{35}(\bissue{6-8}),
\bfpage{591}--\blpage{598}
(\byear{2004})
\end{barticle}
\endbibitem

\bibitem[\protect\citeauthoryear{Huang et~al.}{2024}]{huang2024}
\begin{barticle}
\bauthor{\bsnm{Huang}, \binits{W.Z.}},
\bauthor{\bsnm{Li}, \binits{B.}},
\bauthor{\bsnm{Feng}, \binits{X.Q.}}:
\batitle{Mechanobiological tortuosity of blood vessels with stress-modulated growth and remodeling}.
\bjtitle{Journal of the Mechanics and Physics of Solids}
\bvolume{186},
\bfpage{105605}
(\byear{2024})
\end{barticle}
\endbibitem

\bibitem[\protect\citeauthoryear{Heeman et~al.}{2019}]{heeman2019}
\begin{barticle}
\bauthor{\bsnm{Heeman}, \binits{W.}},
\bauthor{\bsnm{Steenbergen}, \binits{W.}},
\bauthor{\bsnm{Dam}, \binits{G.M.}},
\bauthor{\bsnm{Boerma}, \binits{E.C.}}:
\batitle{Clinical applications of laser speckle contrast imaging: a review}.
\bjtitle{Journal of Biomedical Optics}
\bvolume{24}(\bissue{8}),
\bfpage{080901}
(\byear{2019})
\end{barticle}
\endbibitem

\bibitem[\protect\citeauthoryear{Kassab and Kassab}{2019}]{kassab2019}
\begin{bchapter}
\bauthor{\bsnm{Kassab}, \binits{G.S.}},
\bauthor{\bsnm{Kassab}, \binits{G.S.}}:
\bctitle{Constitutive models of coronary vasculature}.
In: \bbtitle{Coronary Circulation: Anatomy, Mechanical Properties, and Biomechanics},
pp. \bfpage{173}--\blpage{308}
(\byear{2019})
\end{bchapter}
\endbibitem

\bibitem[\protect\citeauthoryear{Linder-Ganz and Gefen}{2007}]{linder2007}
\begin{barticle}
\bauthor{\bsnm{Linder-Ganz}, \binits{E.}},
\bauthor{\bsnm{Gefen}, \binits{A.}}:
\batitle{The effects of pressure and shear on capillary closure in the microstructure of skeletal muscles}.
\bjtitle{Annals of Biomedical Engineering}
\bvolume{35},
\bfpage{2095}--\blpage{2107}
(\byear{2007})
\end{barticle}
\endbibitem

\bibitem[\protect\citeauthoryear{Langemo et~al.}{2008}]{langemo2008}
\begin{barticle}
\bauthor{\bsnm{Langemo}, \binits{D.}},
\bauthor{\bsnm{Thompson}, \binits{P.}},
\bauthor{\bsnm{Hunter}, \binits{S.}},
\bauthor{\bsnm{Hanson}, \binits{D.}},
\bauthor{\bsnm{Anderson}, \binits{J.}}:
\batitle{Heel pressure ulcers: stand guard}.
\bjtitle{Advances in Skin \& Wound Care}
\bvolume{21}(\bissue{6}),
\bfpage{282}--\blpage{292}
(\byear{2008})
\end{barticle}
\endbibitem

\bibitem[\protect\citeauthoryear{Loiseau et~al.}{2019}]{loiseau2019}
\begin{bchapter}
\bauthor{\bsnm{Loiseau}, \binits{E.}},
\bauthor{\bsnm{Viallat}, \binits{A.}},
\bauthor{\bsnm{Abkarian}, \binits{M.}}:
\bctitle{Blood in flow. basic concepts}.
In: \bbtitle{Dynamics of Blood Cell Suspensions in Microflows},
pp. \bfpage{1}--\blpage{39}.
\bpublisher{CRC Press},
\blocation{Boca Raton, FL}
(\byear{2019})
\end{bchapter}
\endbibitem

\bibitem[\protect\citeauthoryear{Murrant and Fletcher}{2022}]{murrant2022}
\begin{barticle}
\bauthor{\bsnm{Murrant}, \binits{C.L.}},
\bauthor{\bsnm{Fletcher}, \binits{N.M.}}:
\batitle{Capillary communication: the role of capillaries in sensing the tissue environment, coordinating the microvascular, and controlling blood flow}.
\bjtitle{American Journal of Physiology-Heart and Circulatory Physiology}
\bvolume{323}(\bissue{5}),
\bfpage{1019}--\blpage{1036}
(\byear{2022})
\end{barticle}
\endbibitem

\bibitem[\protect\citeauthoryear{Negosanti et~al.}{2012}]{negosanti2012}
\begin{barticle}
\bauthor{\bsnm{Negosanti}, \binits{L.}},
\bauthor{\bsnm{Pinto}, \binits{V.}},
\bauthor{\bsnm{Sgarzani}, \binits{R.}},
\bauthor{\bsnm{Negosanti}, \binits{F.}},
\bauthor{\bsnm{Zannetti}, \binits{G.}},
\bauthor{\bsnm{Cipriani}, \binits{R.}}:
\batitle{World journal of}.
\bjtitle{World}
\bvolume{1}(\bissue{3}),
\bfpage{13}--\blpage{23}
(\byear{2012})
\end{barticle}
\endbibitem

\bibitem[\protect\citeauthoryear{Orusbiev et~al.}{2023}]{orusbiev2023}
\begin{barticle}
\bauthor{\bsnm{Orusbiev}, \binits{A.R.}},
\bauthor{\bsnm{Alunkacheva}, \binits{T.G.}},
\bauthor{\bsnm{Charandaeva}, \binits{M.S.}},
\bauthor{\bsnm{Kireeva}, \binits{B.S.}},
\bauthor{\bsnm{Gadzhiev}, \binits{M.F.}},
\bauthor{\bsnm{Zelenetckii}, \binits{V.G.}}:
\batitle{Study of the structural and mechanical properties of erythrocyte membranes using atomic force microscopy}.
\bjtitle{Archives of Pharmacy Practice}
\bvolume{14}(\bissue{2-2023}),
\bfpage{70}--\blpage{74}
(\byear{2023})
\end{barticle}
\endbibitem

\bibitem[\protect\citeauthoryear{Panula}{2022}]{panula2022}
\begin{botherref}
\oauthor{\bsnm{Panula}, \binits{T.}}:
Blood pressure and hemodynamic monitoring from the fingertip.
Master's thesis,
Aalto University
(2022)
\end{botherref}
\endbibitem

\bibitem[\protect\citeauthoryear{Peng et~al.}{2019}]{peng2019}
\begin{barticle}
\bauthor{\bsnm{Peng}, \binits{Z.}},
\bauthor{\bsnm{Shu}, \binits{B.}},
\bauthor{\bsnm{Zhang}, \binits{Y.}},
\bauthor{\bsnm{Wang}, \binits{M.}}:
\batitle{Endothelial response to pathophysiological stress}.
\bjtitle{Arteriosclerosis, Thrombosis, and Vascular Biology}
\bvolume{39}(\bissue{11}),
\bfpage{233}--\blpage{243}
(\byear{2019})
\end{barticle}
\endbibitem

\bibitem[\protect\citeauthoryear{Rastel and Lun}{2019}]{rastel2019}
\begin{barticle}
\bauthor{\bsnm{Rastel}, \binits{D.}},
\bauthor{\bsnm{Lun}, \binits{B.}}:
\batitle{Lower limb deep vein diameters beneath medical compression stockings in the standing position}.
\bjtitle{European Journal of Vascular and Endovascular Surgery}
\bvolume{57}(\bissue{2}),
\bfpage{276}--\blpage{282}
(\byear{2019})
\end{barticle}
\endbibitem

\bibitem[\protect\citeauthoryear{Roy and Secomb}{2022}]{roy2022}
\begin{barticle}
\bauthor{\bsnm{Roy}, \binits{T.K.}},
\bauthor{\bsnm{Secomb}, \binits{T.W.}}:
\batitle{Functional implications of microvascular heterogeneity for oxygen uptake and utilization}.
\bjtitle{Physiological Reports}
\bvolume{10}(\bissue{10}),
\bfpage{15303}
(\byear{2022})
\end{barticle}
\endbibitem

\bibitem[\protect\citeauthoryear{Sloop et~al.}{2020}]{sloop2020}
\begin{botherref}
\oauthor{\bsnm{Sloop}, \binits{G.D.}},
\oauthor{\bsnm{De~Mast}, \binits{Q.}},
\oauthor{\bsnm{Pop}, \binits{G.}},
\oauthor{\bsnm{Weidman}, \binits{J.J.}},
\oauthor{\bsnm{Cyr}, \binits{J.A.S.}}:
The role of blood viscosity in infectious diseases.
Cureus
\textbf{12}(2)
(2020)
\end{botherref}
\endbibitem

\bibitem[\protect\citeauthoryear{Sharzehee et~al.}{2019}]{sharzehee2019}
\begin{barticle}
\bauthor{\bsnm{Sharzehee}, \binits{M.}},
\bauthor{\bsnm{Fatemifar}, \binits{F.}},
\bauthor{\bsnm{Han}, \binits{H.C.}}:
\batitle{Computational simulations of the helical buckling behavior of blood vessels}.
\bjtitle{International Journal for Numerical Methods in Biomedical Engineering}
\bvolume{35}(\bissue{12}),
\bfpage{3277}
(\byear{2019})
\end{barticle}
\endbibitem

\bibitem[\protect\citeauthoryear{Seddighi and Han}{2021}]{seddighi2021}
\begin{barticle}
\bauthor{\bsnm{Seddighi}, \binits{Y.}},
\bauthor{\bsnm{Han}, \binits{H.C.}}:
\batitle{Buckling of arteries with noncircular cross sections: theory and finite element simulations}.
\bjtitle{Frontiers in Physiology}
\bvolume{12},
\bfpage{712636}
(\byear{2021})
\end{barticle}
\endbibitem

\bibitem[\protect\citeauthoryear{Stompor and Perkowska-Ptasińska}{2020}]{stompor2020}
\begin{barticle}
\bauthor{\bsnm{Stompor}, \binits{T.}},
\bauthor{\bsnm{Perkowska-Ptasińska}, \binits{A.}}:
\batitle{Hypertensive kidney disease: a true epidemic or rare disease}.
\bjtitle{Polish Archives of Internal Medicine}
\bvolume{130}(\bissue{2}),
\bfpage{130}--\blpage{139}
(\byear{2020})
\end{barticle}
\endbibitem

\bibitem[\protect\citeauthoryear{Sree et~al.}{2019}]{sree2019}
\begin{barticle}
\bauthor{\bsnm{Sree}, \binits{V.D.}},
\bauthor{\bsnm{Rausch}, \binits{M.K.}},
\bauthor{\bsnm{Tepole}, \binits{A.B.}}:
\batitle{Linking microvascular collapse to tissue hypoxia in a multiscale model of pressure ulcer initiation}.
\bjtitle{Biomechanics and Modeling in Mechanobiology}
\bvolume{18},
\bfpage{1947}--\blpage{1964}
(\byear{2019})
\end{barticle}
\endbibitem

\bibitem[\protect\citeauthoryear{Song et~al.}{2024}]{song2024}
\begin{barticle}
\bauthor{\bsnm{Song}, \binits{B.}},
\bauthor{\bsnm{Wang}, \binits{C.}},
\bauthor{\bsnm{Fan}, \binits{S.}},
\bauthor{\bsnm{Zhang}, \binits{L.}},
\bauthor{\bsnm{Zhang}, \binits{C.}},
\bauthor{\bsnm{Xiong}, \binits{W.}},
\bauthor{\bsnm{Li}, \binits{J.}}:
\batitle{Rapid construction of 3d biomimetic capillary networks with complex morphology using dynamic holographic processing}.
\bjtitle{Advanced Functional Materials}
\bvolume{34}(\bissue{1}),
\bfpage{2305245}
(\byear{2024})
\end{barticle}
\endbibitem

\bibitem[\protect\citeauthoryear{Sun et~al.}{2023}]{sun2023}
\begin{barticle}
\bauthor{\bsnm{Sun}, \binits{W.K.}},
\bauthor{\bsnm{Yin}, \binits{B.B.}},
\bauthor{\bsnm{Zhang}, \binits{L.W.}},
\bauthor{\bsnm{Liew}, \binits{K.M.}}:
\batitle{Blood pressure-driven rupture of blood vessels}.
\bjtitle{Journal of the Mechanics and Physics of Solids}
\bvolume{174},
\bfpage{105274}
(\byear{2023})
\end{barticle}
\endbibitem

\bibitem[\protect\citeauthoryear{Tagliabue}{2023}]{tagliabue2023}
\begin{barticle}
\bauthor{\bsnm{Tagliabue}, \binits{S.e.a.}}:
\batitle{Microvascular cerebral blood flow dynamics for intracranial pressure estimates: transcranial diffuse correlation spectroscopy}.
\bjtitle{Metabolism}
\bvolume{43}(\bissue{1S}),
\bfpage{84}--\blpage{197}
(\byear{2023})
\end{barticle}
\endbibitem

\bibitem[\protect\citeauthoryear{Thiriet}{2019}]{thiriet2019}
\begin{barticle}
\bauthor{\bsnm{Thiriet}, \binits{M.}}:
\batitle{Input data for computational models of blood flows}.
\bjtitle{Digital Medicine}
\bvolume{5}(\bissue{4}),
\bfpage{141}--\blpage{153}
(\byear{2019})
\end{barticle}
\endbibitem

\bibitem[\protect\citeauthoryear{Tooke and Smaje}{2019}]{tooke2019microcirculation}
\begin{bchapter}
\bauthor{\bsnm{Tooke}, \binits{J.E.}},
\bauthor{\bsnm{Smaje}, \binits{L.H.}}:
\bctitle{The microcirculation and clinical disease}.
In: \bbtitle{Clinically Applied Microcirculation Research},
pp. \bfpage{3}--\blpage{16}.
\bpublisher{Routledge},
\blocation{London}
(\byear{2019})
\end{bchapter}
\endbibitem

\bibitem[\protect\citeauthoryear{Tong et~al.}{2019}]{tong2019}
\begin{bchapter}
\bauthor{\bsnm{Tong}, \binits{L.S.}},
\bauthor{\bsnm{Yu}, \binits{Y.N.}},
\bauthor{\bsnm{Tang}, \binits{J.}},
\bauthor{\bsnm{Lou}, \binits{M.}},
\bauthor{\bsnm{Zhang}, \binits{J.H.}}:
\bctitle{Involvement of cerebral venous system in ischemic stroke}.
In: \bbtitle{Cerebral Venous System in Acute and Chronic Brain Injuries},
pp. \bfpage{195}--\blpage{205}
(\byear{2019})
\end{bchapter}
\endbibitem

\bibitem[\protect\citeauthoryear{Wanless}{2020}]{wanless2020}
\begin{barticle}
\bauthor{\bsnm{Wanless}, \binits{I.R.}}:
\batitle{The role of vascular injury and congestion in the pathogenesis of cirrhosis: the congestive escalator and the parenchymal extinction sequence}.
\bjtitle{Current Hepatology Reports}
\bvolume{19},
\bfpage{40}--\blpage{53}
(\byear{2020})
\end{barticle}
\endbibitem

\end{thebibliography}

\end{document}